# Fermi surface reconstruction in electron-doped cuprates without antiferromagnetic long-range order

J.-F. He[1,2], C. R. Rotundu[1,2], M. S. Scheurer[3], Y. He[1,2], M. Hashimoto[4], K. Xu[2], Y. Wang[1,3], E. W. Huang[1,2], T. Jia[2], S.-D. Chen[1,2], B. Moritz[1,2], D.-H. Lu[4], Y. S. Lee[1,2], T. P. Devereaux[1,2] & Z.-X. Shen[1,2,*]

[1]Stanford Institute for Materials and Energy Sciences, SLAC National Accelerator Laboratory, 2575 Sand Hill Road, Menlo Park, California 94025, USA

[2]Geballe Laboratory for Advanced Materials, Departments of Physics and Applied Physics, Stanford University, Stanford, California 94305, USA

[3]Department of Physics, Harvard University, Cambridge MA 02138, USA

[4]Stanford Synchrotron Radiation Lightsource, SLAC National Accelerator Laboratory, 2575 Sand Hill Road, Menlo Park, California 94025, USA

*Correspondence and requests for materials should be addressed to Z.X.S. (zxshen@stanford.edu)

**Abstract**

Fermi surface (FS) topology is a fundamental property of metals and superconductors. In electron-doped cuprate $Nd_{2-x}Ce_xCuO_4$ (NCCO), an unexpected FS reconstruction has been observed in optimal- and over-doped regime (x=0.15-0.17) by quantum oscillation measurements (QOM). This is all the more puzzling because neutron scattering suggests that the antiferromagnetic (AFM) long-range order, which is believed to reconstruct the FS, vanishes before x=0.14. To reconcile the conflict, a widely discussed external magnetic field-induced AFM long-range order in QOM explains the FS reconstruction as an extrinsic property. Here, we report angle-resolved photoemission (ARPES) evidence of FS reconstruction in optimal- and over-doped NCCO. The observed FSs are in quantitative agreement with QOM, suggesting an intrinsic FS reconstruction without field. This reconstructed FS, despite its importance as a basis to understand electron-doped cuprates, cannot be explained under the traditional scheme. Furthermore, the energy gap of the reconstruction decreases rapidly near x=0.17 like an order parameter, echoing the quantum critical doping in transport. The totality of the data points to a mysterious order between x=0.14 and 0.17, whose appearance favors the FS reconstruction and disappearance defines the quantum critical doping. A recent topological proposal provides an ansatz for its origin.

Fermi surface topology is the starting point to understand various emergent quantum phenomena in metals, including high-temperature superconductivity. With both momentum and energy resolution, ARPES is an ideal tool to directly reveal the FS topology of a material. However, in electron-doped cuprates, a direct understanding of the ARPES results has been limited by the data quality (1-4). This is primarily due to the lack of a large high-quality surface area in a material that is difficult to cleave. Utilizing a newly developed ARPES beam line at Stanford Synchrotron Radiation Lightsource with a small beam spot, we have managed to probe intrinsic electronic structures from a small but uniform region on the cleaved sample surface. This technical advancement leads to a significant improvement on the experimental data quality (see supplementary Fig. 1) that enables us to quantitatively investigate the FS topology in electron-doped cuprates.

When a FS reconstruction takes place, the energy band is folded with respect to the antiferromagnetic zone boundary (AFMZB) and an energy gap opens up, giving rise to a back-bending behavior of the band at the AFMZB (5) (see Fig. 1A-C for a schematic diagram). If the gap is below Fermi level ($E_F$) (Fig. 1B), then the $E_F$ cuts through the conduction band, resulting in an electron-like pocket (e.g. antinodal region in Fig. 1A). Conversely, if the gap is above $E_F$ (Fig. 1C), a hole-like pocket appears on the FS (e.g. nodal region in Fig. 1A). On the other hand, when the FS reconstruction is absent (Fig. 1D), the electron band disperses continuously, irrespective of the AFMZB (Fig. 1E,1F). Neither band folding nor gap opening is expected.

Earlier ARPES measurements on underdoped samples have revealed the AFM gap (2-7), hints of the folded band (8) and disconnected segments on the FS (2-7), supporting the reconstruction scenario in underdoped regime (2). However, things become more complicated with electron doping (4,6,9-11). Photoemission constant energy map at $E_F$ of the optimal-doped $Nd_{2-x}Ce_xCuO_4$ (NCCO, x=0.15) seems to suggest a large FS centered at ($\pi$, $\pi$) (4). But a spectral weight analysis of the nodal dispersion favors a reconstructed FS for the optimal-doped $Sm_{2-x}Ce_xCuO_{4-\delta}$ (SCCO, x=0.15) (10). While slight variations between different material families have been discussed (11), a direct understanding of the FS topology requires a better resolution of the key features – band folding and gap opening at the AFMZB.

**Results**

With the improved precision of data, our measurements on the optimal-doped NCCO (x=0.15) clearly reveal both band folding and gap opening at the AFMZB (momentum cut near the "hotspot" where the FS intersects AFMZB, Fig. 1G-I), demonstrating the existence of a reconstructed electron-like pocket. The FS reconstruction is also supported by the nodal dispersion (Fig. 1J-L), where the Fermi level crosses the hole band, forming a hole-like pocket near the node. The possible gap opening above $E_F$ cannot be seen by ARPES, but the back-bending of the hole band (folded band) is discernible (Fig. 1J-L). A reconstructed FS should also be accompanied by a hotspot between the electron-like and hole-like pockets, where a gap exists at the Fermi level (Fig. 1A). This is also observed in our experiment (see supplementary Fig. 2 for the FS mapping and the momentum cut through the hotspot).

After establishing the FS reconstruction in optimal-doped NCCO, we quantitatively investigate the associated energy gap. Surprisingly, the gap shows a strong momentum dependence. A ~80 meV gap appears near the hotspot (Fig. 2B), but it vanishes at the AFMZB near the antinode (Fig. 2C, 2H and supplementary Fig. 3). This is distinct from the mean field band-folding picture, where a momentum independent constant energy gap is expected (Fig. 2E-G). To understand the differences, we note that a moderate energy gap can be smeared out on the photoelectron spectra when an enhanced scattering rate takes place (see supplementary Fig. 4). The measured electron scattering rate, represented by the width of the momentum distribution curves (MDC) near $E_F$ (Fig. 2I), does show a substantial momentum dependence (Fig. 2J). The enhanced scattering rate near antinode, when combined with its deep binding energy where the band cross the AFMZB, could give rise to the folded band without a resolvable gap opening (see supplementary Fig. 4). As such, the gap itself is likely isotropic and the apparent momentum dependence of the gap in Fig. 2H can be attributed to the scattering rate difference. Although the absolute value of scattering rate deduced from ARPES is different from that in transport, the strong momentum dependence coincides with the fact that only the hole-like pocket near node has been observed in QOM, while the expected electron-like pocket near antinode is absent (12-17). The origin of the momentum dependent scattering rate is yet to be understood, where the electron correlation might be at play.

Next, we study the doping dependence of the FS reconstruction. Both the back-bending behavior (folded band) and gap opening have been observed at all doping levels we measured (x=0.11, x=0.15,

x=0.16), demonstrating that the FS reconstruction persists to the over-doped regime. This is consistent with QOM (12-17), and the well-defined gap suggests that a magnetic breakdown in QOM is less likely below x=0.16. However, the gap decreases rapidly near x=0.16 (Fig. 3F), which is consistent with our inability to observe a gap beyond x=0.16. Such a behavior of gap closing near x=0.16~0.17 is consistent with a quantum critical point (QCP) in that doping range, as suggested by a x to 1-x density transition in transport experiments (13,18,19). The size of the reconstructed Fermi pockets can also be measured. One way is to directly estimate the area of the pockets via FS mapping (Fig. 3E and supplementary Fig. 5). Alternatively, one can also estimate the pocket size using the reconstructed band dispersion and the gap size (see supplementary for details). Both methods give similar results, which are quantitatively consistent with quantum oscillation results as in Fig. 3G. The similar Fermi pockets observed by our measurements and QOM indicate the existence of a robust FS reconstruction, regardless of the external magnetic field.

**Discussion**

One scenario for the ARPES data is a remnant short-range order (8,20,21,22). Strictly speaking, a short-range order does not break the global translational symmetry of the crystal. However, it might provide a scattering channel with a wave vector Q±ΔQ, where ΔQ is proportional to the inverse of the correlation length, and thus an approximate "FS folding". Nevertheless, with the small AFM correlation length in over-doped NCCO (x=0.16) of ~7a (23), a weak coupling mean-field simulation does not capture the experimental observations (see supplementary Fig. 6). Our data, on the other hand, leave room for a strong coupling picture with short-range AFM fluctuation, as those suggested by the Hubbard model calculations. Here the momentum folding remains commensurate with the local correlation at Q = ($\pi$, $\pi$), and the gap magnitude is also dominated by the local interaction and thus remains similar in different regions (see supplementary Fig. 7). Our experimental energy and momentum width allow such a picture. However, it is unclear whether the quantum critical doping, as seen by the rapid decrease of the energy gap near x=0.16 (Fig. 3F) and the corresponding transport data (13,18,19,24-26), can be understood by such a purely local picture.

The totality of ARPES, QOM, and transport data suggests the presence of an intrinsic long-range order that persists up to the critical doping near x=0.16-0.17. Long-range AFM order can naturally explain this phenomenology. However, neutron scattering data from the rod-like magnetic scattering in the bulk indicate the lack of co-existence between long-range AFM order and superconductivity beyond x=0.14 (23). A similar conclusion is drawn by the µSR measurements on another electron-doped cuprate $La_{2-x}Ce_xCuO_{4-\delta}$, where the static magnetism and superconductivity do not coexist (27). One is therefore left with a puzzle on the origin of the long-range order in our superconducting NCCO samples with x=0.15-0.16 doping.

Charge order has been reported in NCCO (28), but the associated wave vector doesn't match the observed FS reconstruction. Another possible way out involves topological order that exploits the topological character of Luttinger theorem (29-31). Without breaking the translational symmetry, the existence of topological order in a state with short-range AFM order can still reconstruct the FS with respect to the AFMZB (29-31) (also see supplementary Fig. 8). In the SU(2) gauge theory, the gauge-dependent Higgs field cannot be directly observed, but can play a role similar to an order parameter. Its presence could have observable consequences like the opening of the gap, which has the meaning of the magnitude of the local magnetic order and its closing defines a QCP. It would be instructive to have deeper understanding of the transport behavior near a topological QCP for refined comparison with experiment. It would also be interesting to explore whether the same basic scenario can be at play in hole-doped cuprates.

Through much improved experiment, our data have established the intrinsic doping dependence of FS topology in NCCO, and provided a microscopic underpinning for QOM without the need to assume magnetic field induced long-range AFM order in the optimal- and over-doped regime. The rapid closing of the gap near x=0.16-0.17 uncovers the microscopic origin of the quantum critical doping in transport experiments (13,18,19,24-26). Confronted by the neutron conclusion of an absence of long-range AFM order beyond x=0.14 (23), a correlation driven topological order provides an ansatz to reconcile the dilemma.

## Materials and Methods

**Samples.** Single crystals of NCCO (x=0.11, 0.15 and 0.16) were grown by the traveling-solvent floating-zone method in $O_2$ and annealed in Ar. The doping levels were determined by electron probe microanalysis (EPMA).

**ARPES.** ARPES measurements were carried out at beam line 5-2 of the Stanford Synchrotron Radiation Lightsource of SLAC National Accelerator Laboratory with a total energy resolution of ~12 meV and a base pressure better than $5 \times 10^{-11}$ Torr. The data were collected with 53eV photons at ~20 K. The Fermi level was referenced to that of a polycrystalline Au film in electrical contact with the sample. The smallest beam spot size at the beam line is ~40 μm (horizontal) X 10 μm (vertical). For our experiments, a beam spot of ~40 μm (horizontal) X 80 μm (vertical) was chosen, which optimized the photoelectron counts on a single uniform surface region.

**ACKNOWLEDGMENTS.** We thank S. Sachdev, D. J. Scalapino, D.-H. Lee, P. J. Hirschfeld, S. A. Kivelson, J. Zaanen, N. Nagaosa, A. Georges, C. M. Varma and S. Uchida for useful discussions. The work at SLAC and Stanford is supported by the US DOE, Office of Basic Energy Science, Division of Materials Science and Engineering. SSRL is operated by the Office of Basic Energy Sciences, US DOE, under contract No. DE-AC02-76SF00515. M. S. S. acknowledges support from the German National Academy of Sciences Leopoldina through Grant No. LPDS 2016-12.

1. Damascelli A, Hussain Z, Shen ZX (2003) Angle-resolved photoemission studies of the cuprate superconductors. *Rev Mod Phys* 75:473-541.

2. Armitage NP, Fournier P, Greene RL (2010) Progress and perspectives on electron-doped cuprates. *Rev Mod Phys* 82:2421-2487.

3. Armitage NP (2001) Doping the copper-oxygen planes with electrons: the view with photoemission. Ph.D. thesis (Stanford University).

4. Armitage NP, et al. (2002) Doping dependence of an n-type cuprate superconductor investigated by angle-resolved photoemission spectroscopy. *Phys Rev Lett* 88:257001.

5. Matsui H, et al. (2005) Angle-resolved photoemission spectroscopy of the antiferromagnetic superconductor $Nd_{1.87}Ce_{0.13}CuO_4$: anisotropic spin-correlation gap, pseudogap, and the induced quasiparticle mass enhancement. *Phys Rev Lett* 94:047005.

6. Matsui H, et al. (2007) Evolution of the pseudogap across the magnet-superconductor phase boundary of $Nd_{2-x}Ce_xCuO_4$. *Phys Rev B* 75:224514.

7. Matsui H, et al. (2005) Direct observation of a nonmonotonic $d_{x^2-y^2}$-wave superconducting gap in the electron-doped high-Tc superconductor $Pr_{0.89}LaCe_{0.11}CuO_4$. *Phys Rev Lett* 95:017003.

8. Park SR, et al. (2007) Electronic structure of electron-doped $Sm_{1.86}Ce_{0.14}CuO_4$: Strong pseudogap effects, nodeless gap, and signatures of short-range order. *Phys Rev B* 75:060501(R).

9. Armitage NP, et al. (2001) Anomalous electronic structure and pseudogap effects in $Nd_{1.85}Ce_{0.15}CuO_4$. *Phys Rev Lett* 87:147003.

10. Santander-Syro AF, et al. (2011) Two-fermi-surface superconducting state and a nodal d-wave energy gap of the electron-doped $Sm_{1.85}Ce_{0.15}CuO_{4-\delta}$ cuprate superconductor. *Phys Rev Lett* 106:197002.

11. Ikeda M, et al. (2009) Effects of chemical pressure on the Fermi surface and band dispersion of the electron-doped high-$T_c$ superconductors. *Phys Rev B* 80:014510.

12. Helm T, et al. (2009) Evolution of the Fermi surface of the electron-doped high-temperature superconductor $Nd_{2-x}Ce_xCuO_4$ revealed by Shubnikov–de Haas oscillations. *Phys Rev Lett* 103: 157002.

13. Helm T, et al. (2010) Magnetic breakdown in the electron-doped cuprate superconductor $Nd_{2-x}Ce_xCuO_4$: The reconstructed Fermi surface survives in the strongly overdoped regime. *Phys Rev Lett* 105:247002.

14. Kartsovnik MV, et al. (2011) Fermi surface of the electron-doped cuprate superconductor $Nd_{2-x}Ce_xCuO_4$ probed by high-field magnetotransport. *New Journal of Physics* 13:015001.

15. Helm T, et al. (2015) Correlation between Fermi surface transformations and superconductivity in the electron-doped high-Tc superconductor $Nd_{2-x}Ce_xCuO_4$. *Phys Rev B* 92:094501.

16. Breznay NP, et al. (2016) Shubnikov-de Haas quantum oscillations reveal a reconstructed Fermi surface near optimal doping in a thin film of the cuprate superconductor $Pr_{1.86}Ce_{0.14}CuO_{4\pm\delta}$. *Phys Rev B* 94:104514.

17. Higgins JS, et al. (2018) Quantum oscillations from the reconstructed Fermi surface in electron-doped cuprate superconductors. *arXiv:* 1804.06317.

18. Dagan Y, Qazilbash MM, Hill CP, Kulkarni VN, Greene RL (2004) Evidence for a quantum phase transition in $Pr_{2-x}Ce_xCuO_{4-\delta}$ from transport measurements. *Phys Rev Lett* 92:167001.

19. Zhang X, et al. (2016) Transport anomalies and quantum criticality in electron-doped cuprate superconductors. *Physica C: Superconductivity and its applications* 525-526:18–43.

20. Song D, et al. (2017) Electron number-based phase diagram of $Pr_{1-x}LaCe_xCuO_{4-\delta}$ and possible absence of disparity between electron- and hole-doped cuprate phase diagrams. Phys Rev Lett 118:137001.

21. Horio M, et al. (2018) Common origin of the pseudogap in electron-doped and hole-doped cuprates governed by Mott physics. *arXiv:* 1801.04247.

22. Yu W, Higgins JS, Bach P, Greene RL (2007) Transport evidence of a magnetic quantum phase transition in electron-doped high-temperature superconductors. *Phys Rev B 76: 020503 (R).*

23. Motoyama EM, et al. (2007) Spin correlations in the electron-doped high-transition-temperature superconductor $Nd_{2-x}Ce_xCuO_{4\pm\delta}$. *Nature* 445:186-189.

24. Jin K, Butch NP, Kirshenbaum K, Paglione J, Greene RL (2011) Link between spin fluctuations and electron pairing in copper oxide superconductors. *Nature* 476:73-75.

25. Butch NP, Jin K, Kirshenbaum K, Greene RL, Paglione J (2012) Quantum critical scaling at the edge of Fermi liquid stability in a cuprate superconductor. *Proc Nat Acad Sci* 109:8440-8444.

26. Tafti FF, et al. (2014) Nernst effect in the electron-doped cuprate superconductor $Pr_{2-x}Ce_xCuO_4$: Superconducting fluctuations, upper critical field $H_{c2}$, and the origin of the $T_c$ dome. *Phys Rev B* 90:024519.

27. Saadaoui H, et al. (2015) The phase diagram of electron-doped $La_{2-x}Ce_xCuO_{4-\delta}$. *Nat Commun* 6:6041.

28. Neto EH da S, et al. (2015) Charge ordering in the electron-doped superconductor $Nd_{2-x}Ce_xCuO_4$. *Science* 347:282–285.

29. Scheurer MS, et al. (2018) Topological order in the pseudogap metal. *Proc Nat Acad Sci* 115:E3665-E3672.

30. Sachdev S (2018) Topological order and Fermi surface reconstruction. *arXiv:* 1801.01125v3.

31. Wu W, et al. (2018) Pseudogap and Fermi-surface topology in the two-dimensional Hubbard model. *Phys Rev X* 8:021048.

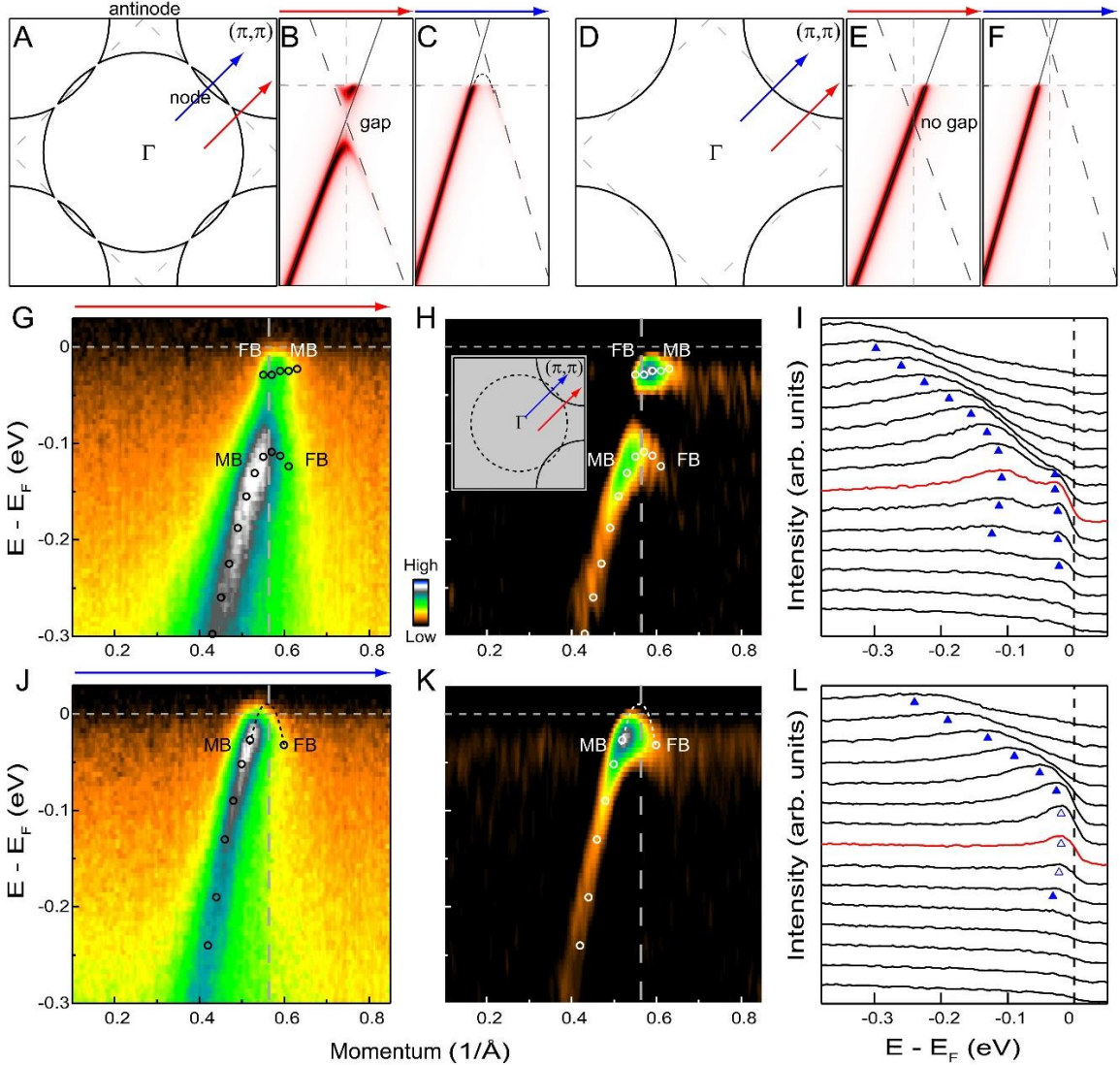

*Fig. 1. Fermi surface reconstruction in optimal-doped NCCO.* (A) Schematic diagram of a reconstructed FS with electron-like pockets near antinode and hole-like pockets near node. The dashed lines indicate the antiferromagnetic Brillouin zone. (B) Schematic band dispersion along a momentum cut on the electron-like pocket (near hotspot), marked by the red arrow in A. The original dispersion is split into conduction and valence bands by an AFM energy gap. The reconstructed bands bend back at the AFMZB. (C) Schematic band dispersion along a momentum cut on the hole-like pocket (nodal cut), marked by the blue arrow in A. The AFM energy gap is slightly above $E_F$, but the folded band (back-bent hole band) disperses below $E_F$. The gray (dashed) line in B and C represents the original (folded) band. (D-F) The same as A-C, but for the original FS without reconstruction. (G-I) Photoemission intensity plot (G), second derivative image with respect to energy (H) and raw energy distribution curves (EDCs) (I) for optimal-doped NCCO, measured along a momentum cut on the electron-like pocket (near hotspot, labelled by the red arrow in the inset of H). Conduction and valence bands extracted from the EDCs (blue triangles in I) are also presented in G (black circles) and H (white circles). The EDC at the AFMZB is shown in red (I). The main band and folded band are marked by "MB" and "FB", respectively. (J-L) The same as G-I, but for the nodal cut (labelled by the blue arrow in the inset of H).

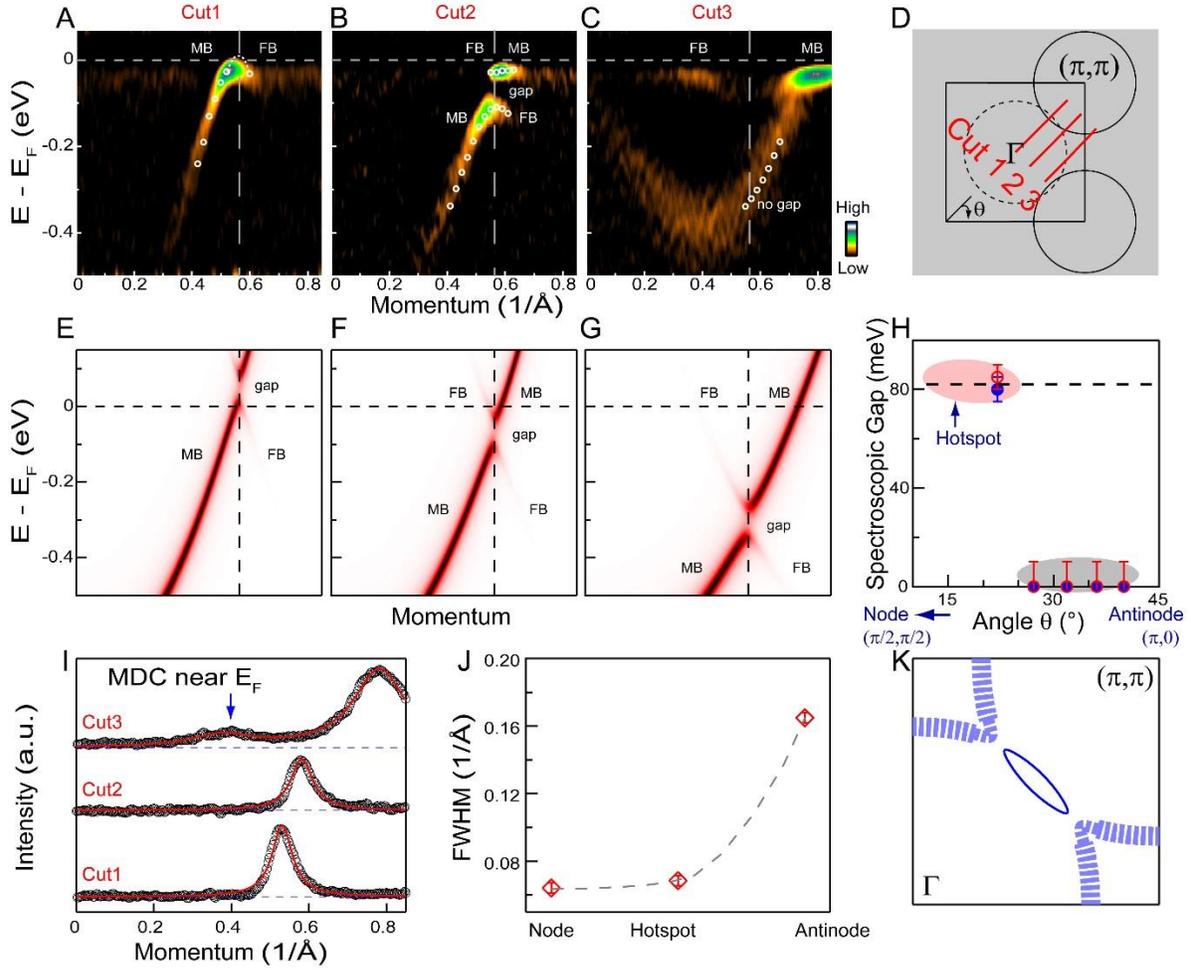

*Fig. 2. Momentum dependent gap in optimal-doped NCCO.* (A-C) Band dispersion (second derivative image with respect to energy) along three momentum cuts as labelled in D. The vertical dashed lines mark the AFMZB. The kink feature (~65 meV) in C comes from electron-boson coupling (see supplementary Fig. 3 for more details). The main band and folded band are marked by "MB" and "FB", respectively. (E-G) Mean-field simulations with a momentum independent gap of 82 meV. The energy of the intersecting point between the original band and the AFMZB for each spectrum is selected to mimic the experiment. (H) Momentum dependence of the measured spectroscopic gap. Note that both intrinsic gap and scattering rate contribute to the spectroscopic gap. We emphasize that the folded conduction band is still clearly observed near the antinode (C, I), forming the reconstructed electron-like Fermi pocket. (I) Integrated momentum distribution curves (MDCs) near $E_F$ (-0.03~0 eV) for the three cuts. The blue arrow marks the MDC peak from the folded band. (J) Full width at half maximum (FWHM) extracted from the main MDC peaks in I. The error bar comes from the fitting. Note that different slopes of the band dispersion at different momentum locations might also contribute to the change of MDC width near $E_F$. However, the overall dispersion near antinode shows a much larger MDC width than that near node and hotspot, indicating an enhanced scattering rate near antinode. (K) Schematic FS. The blurred electron-like pockets indicate the enhanced scattering rate.

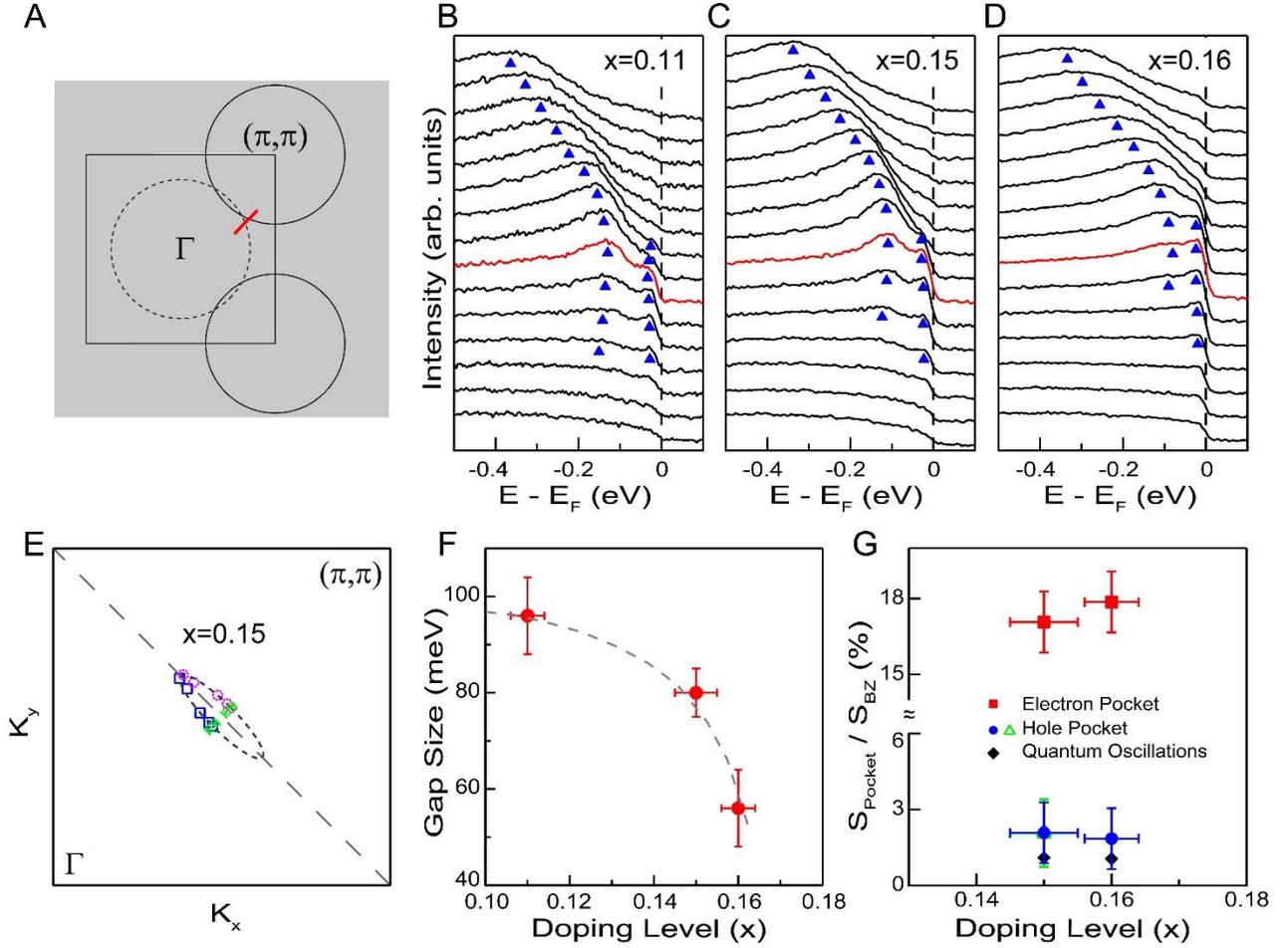

*Fig. 3. Doping dependence of the Fermi surface reconstruction.* (B-D) Raw EDCs for doping levels x=0.11, 0.15 and 0.16 respectively. The corresponding momentum cut is shown in A. The EDCs at the AFMZB are shown in red. Note that both back-bending and gap opening appear right at the AFMZB for all doping levels. These key features differentiate FS reconstruction from bosonic mode coupling (see supplementary Fig. 3 for details). (E) Extracted hole-like Fermi pocket near node. The green squares represent the Fermi momenta ($K_F$) extracted from the main band and folded band along nodal cut. The blue squares mark the $K_F$ obtained from the main band along off-nodal cuts, and the dashed pink circles indicate the "shifted $K_F$" (see supplementary Fig. 5 for details). The error bar comes from the uncertainty in the determination of $K_F$. (F) Doping dependence of the gap size at the AFMZB, estimated by the energy separation between the valence band top and conduction band bottom at the AFMZB, as indicated by the blue triangles on the red EDCs in B-D. Vertical error bars here represent uncertainties of the extracted gap size. Horizontal error bars in F and G represent the uncertainties of the doping levels, determined by electron probe microanalysis (EPMA). (G) Doping dependence of the Fermi pockets. The red squares (electron-like pocket) and blue circles (hole-like pocket) represent the size of the pockets estimated from the energy gap (F). The green triangle (hole-like pocket) shows the value extracted from E. The black diamonds are the results from QOM (13). Vertical error bars represent uncertainties of the pocket size.

## *Supplementary Information*

**Small beam spot and improved data quality**

A small beam spot probes intrinsic electronic structures from a small but uniform region on the terraced sample surface (Fig. 1a). The smallest beam spot size at the new beam line (SSRL BL-52) is ~40 μm (horizontal) X 10 μm (vertical). For our experiments, we chose a beam spot of ~40 μm (horizontal) X 80 μm (vertical) that optimizes the photoelectron counts on a single uniform surface region. The coherent quasi-particle peak (Fig. 1b, enhanced spectral weight near $E_F$) and fine energy features (Fig. 1c,d, dispersion kink) indicate the high data quality.

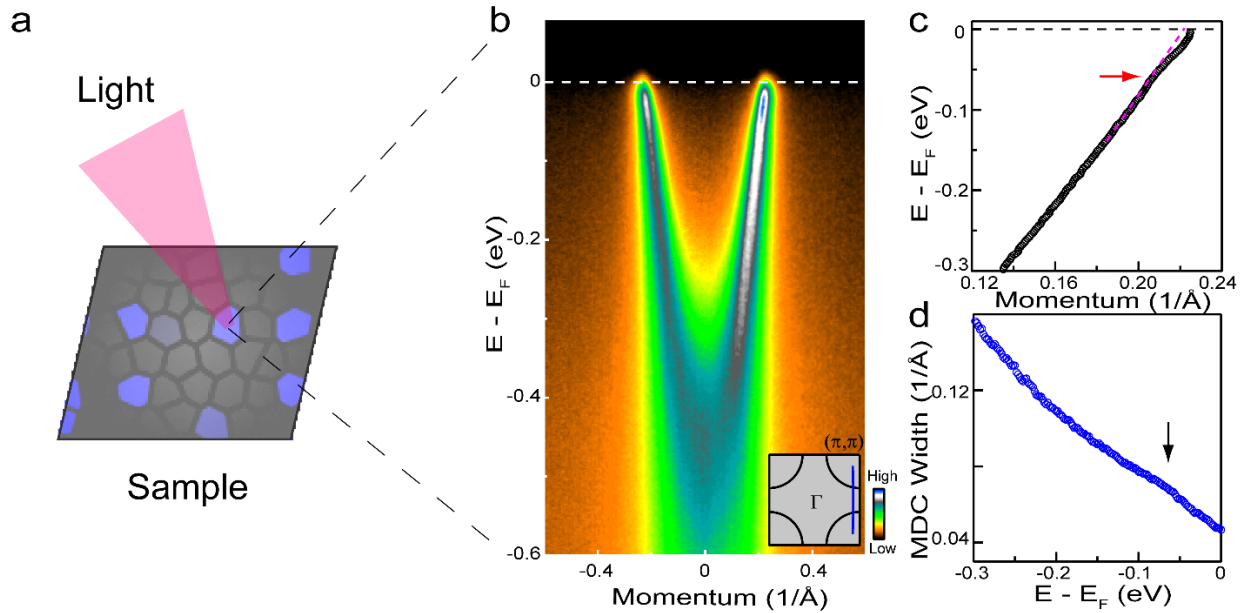

**Fig. 1. Small beam spot and improved data quality. a,** Schematic diagram of the experimental setup. **b,** Photoelectron intensity as a function of energy and momentum. The inset shows the location of the momentum cut. **c**, Band dispersion extracted from the right branch in **b** by fitting MDCs. The red arrow marks the dispersion kink. The pink linear dashed line is a guide to the eye. **d**, The corresponding MDC width as a function of energy. The black arrow marks the drop in MDC width.

**Fermi surface mapping and band dispersion across the hotspot**

The Fermi surface is split into electron-like and hole-like pockets. An energy gap is observed on the band dispersion along a cut through the hotspot, suggesting the reconstruction of the Fermi surface. For this cut, the bottom of the conduction band is above Fermi level ($E_F$). Therefore, no quasiparticle is observed at $E_F$. This is different from a momentum cut close to hotspot but on the electron-like pocket (e.g. cut 2 in main Fig. 2D), where quasiparticles show up near $E_F$, representing the bottom of the electron-like conduction band.

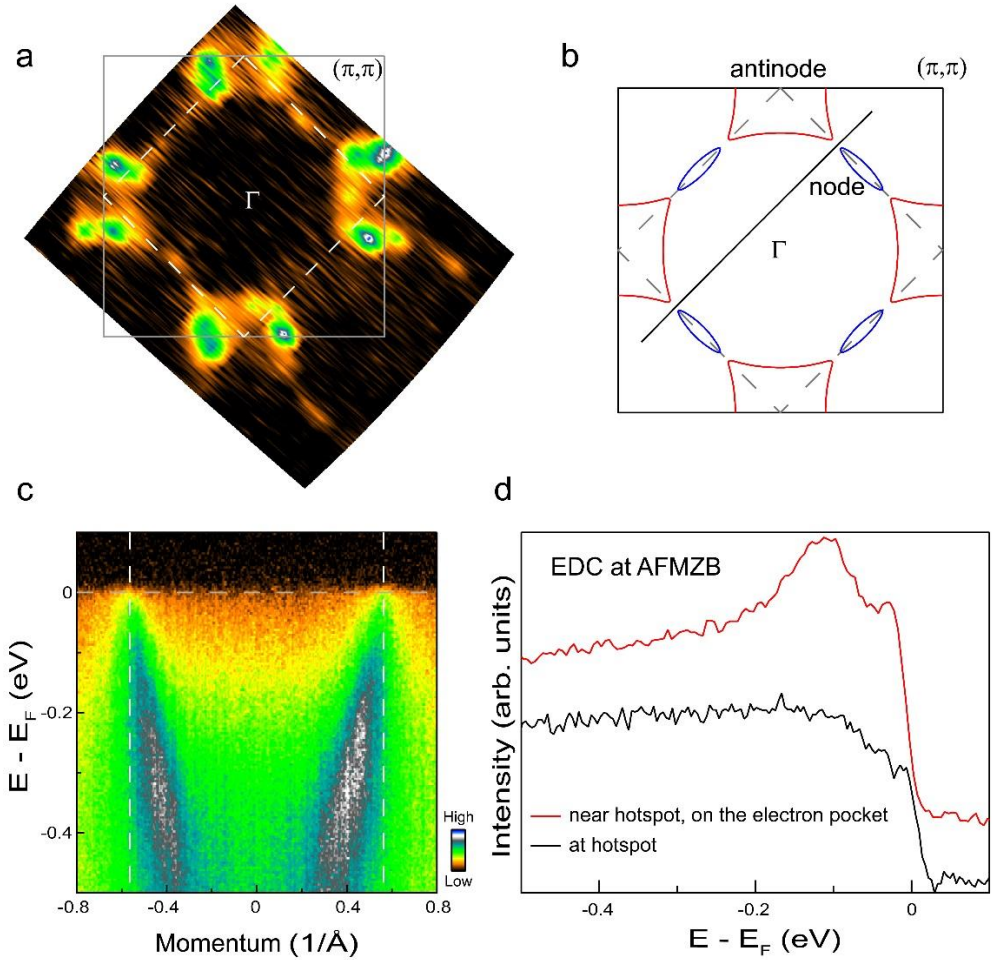

**Fig. 2. Fermi surface mapping and band dispersion across the hotspot (x=0.15). a,** Fermi surface mapping. **c**, Photoemission intensity plot along the momentum cut shown in **b**. **d,** EDCs at the AFMZB. The black curve shows the EDC at the hotspot. The red curve is re-plotted from the red EDC in main Fig. 1I. It is on the AFMZB but slightly away from the hotspot, as shown in main Fig. 1H,I.

**EDC "peak-dip-hump" features: AFM gap vs. electron-boson coupling**

AFM gap, associated with the Fermi surface reconstruction, has been characterized by a distinct depression in electron density of states, reminiscent of a "peak-dip-hump" structure in EDCs of the ARPES spectra. The peak and hump features represent the conduction and valence bands, respectively. Additionally, electron-boson coupling can also give rise to "peak-dip-hump" features. These two phenomena coexist in electron doped cuprates.

Fig. 3 presents band dispersion along two momentum cuts. "Peak-dip-hump" features show up in EDCs of both cuts. However, a closer examination reveals distinct behaviors. First, a back-bending behavior is observed on the dispersion along cut 1, which can be clearly seen on both raw EDCs (Fig. 3c) and the second derivative image (Fig. 3b). This back-bending behavior is absent along cut 2, where both upper and lower branches of the band disperse towards an energy discontinuity asymptotically (Fig. 3e,3f). Second, the minimum energy separation between the peak and hump features along cut 1 locates exactly at the AFMZB, which is not the case for the dispersion along cut 2. These distinctions suggest different origin for the "peak-dip-hump" features along the two momentum cuts. While the characteristic behaviors near the hotspot (cut 1) are consistent with the opening of an AFM gap, those near the antinode (cut 2) represent the hallmarks of a typical electron-boson coupling (the energy discontinuity in the spectrum is the energy of the bosonic mode).

We note that electron-boson coupling has been observed in many cuprates within a wide doping range. Therefore, the EDC "peak-dip-hump" structure alone does not guarantee a Fermi surface reconstruction.

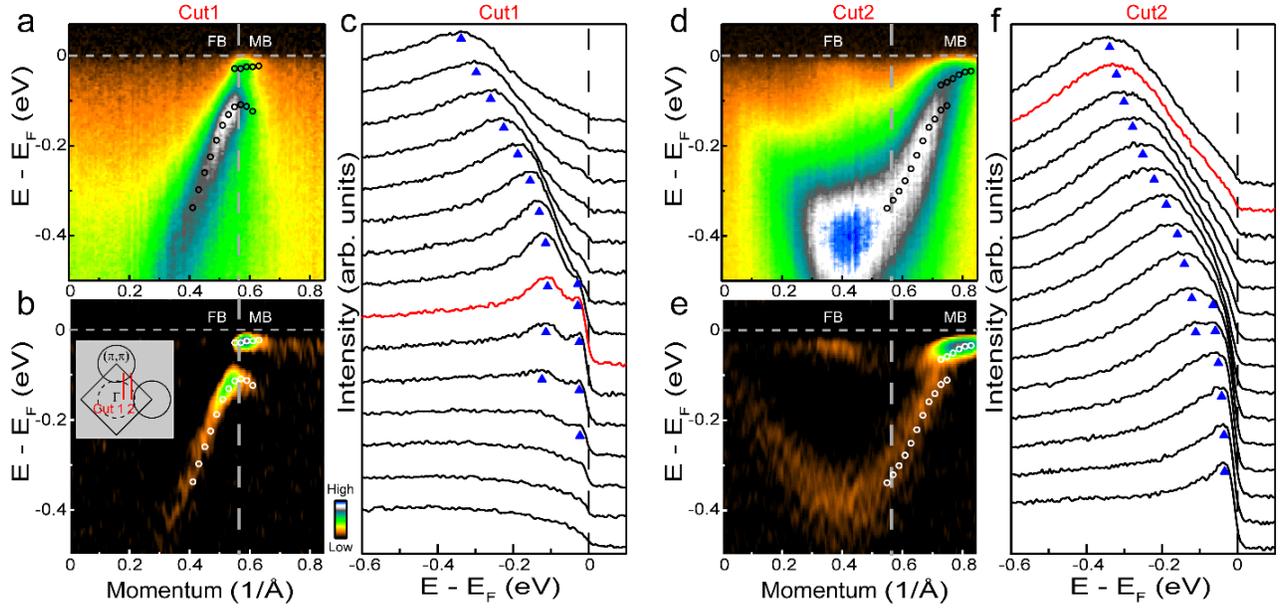

**Fig. 3. Comparison of the electronic structure near the hotspot with that near the antinode. a-c,** Photoelectron intensity plot (**a**), second derivative image with respect to energy (**b**) and raw EDCs (**c**) measured along cut 1. The circles in **a,b** represent the band dispersion extracted from EDC peaks (triangles in **c**). The gray vertical dashed lines in **a,b** label the location of the AFMZB, and the EDC at the AFMZB is shown in red. **d-f,** The same as **a-c,** but for momentum cut 2. The main band and folded band are marked by "MB" and "FB", respectively. The locations of the momentum cuts are shown in the inset of **b.**

**AFM gap smeared out by an increased scattering rate**

To study the effect of scattering rate on the spectroscopic energy gap, electron spectral function has been simulated with two different scattering rates. All the other parameters (bare band, AFM gap size etc.) are the same for both simulations. The AFM gap opening is clearly observed on the spectrum with a low scattering rate (Fig. 4a). This gap is smeared out with an increased scattering rate (Fig. 4b), but the folded band is still discernable near $E_F$ (Fig. 4b).

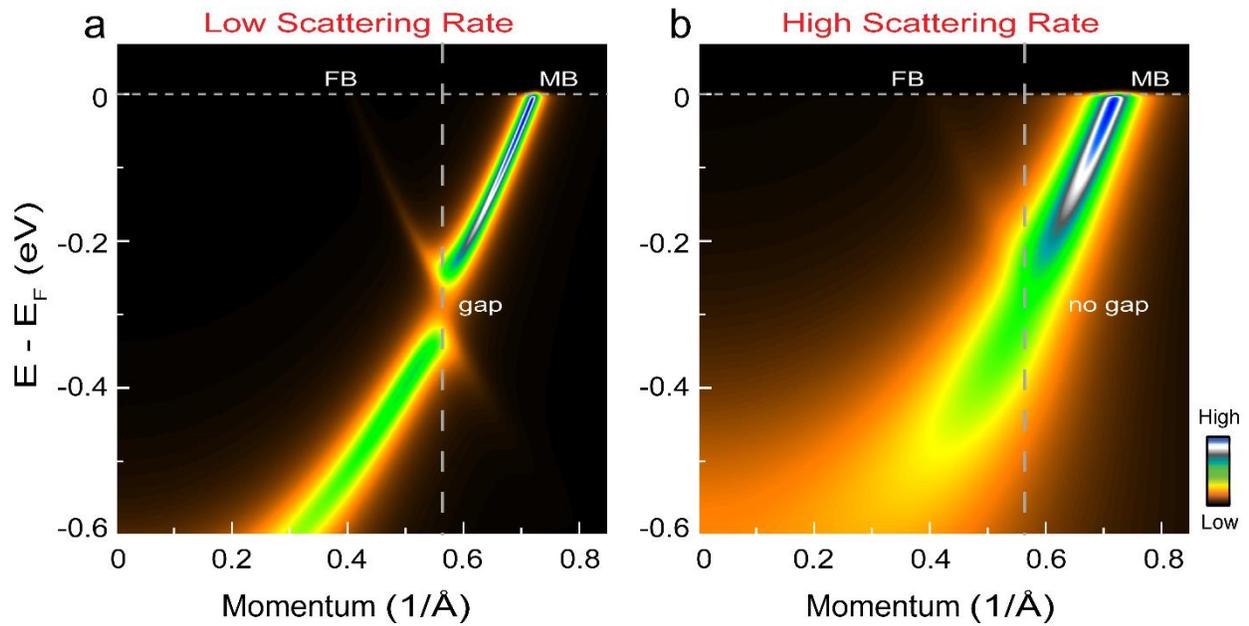

**Fig. 4. Simulated electron spectral function with different scattering rates. a,** Simulated electron spectral function with a low scattering rate. **b,** Simulated electron spectral function with an enhanced scattering rate (close to the experimental scattering rate near antinode). A gap of 100 meV is used for both simulations.

**Extraction of the Fermi momenta for the hole-like pocket.**

In order to quantify the hole-like pocket, we extract the Fermi momentum ($K_F$) of the main band and the folded band respectively. Here, we show the extraction of $K_F$ along the nodal cut. For the main band (e.g. right branch in Fig. 5a), both MDC (red bar) and EDC (blue bar) analyses give the same $K_F$ (marked by the left blue bar in Fig. 5a; the red bar is overlapped by the blue bar and cannot be seen). For the folded band, the intensity is relatively weak. It is hard to determine the $K_F$ by MDC. But EDC analysis still shows the $K_F$ clearly (the lower red arrow in Fig. 5c and the right blue bar in Fig. 5a). Alternatively, we can extract the $K_F$ of the left branch main band (the left red bar in Fig. 5a) and shift it by ($\pi$, $\pi$). This also gives the $K_F$ of the right branch folded band (the pink bar in Fig. 5a). Consistent results are obtained by both methods within the error bar.

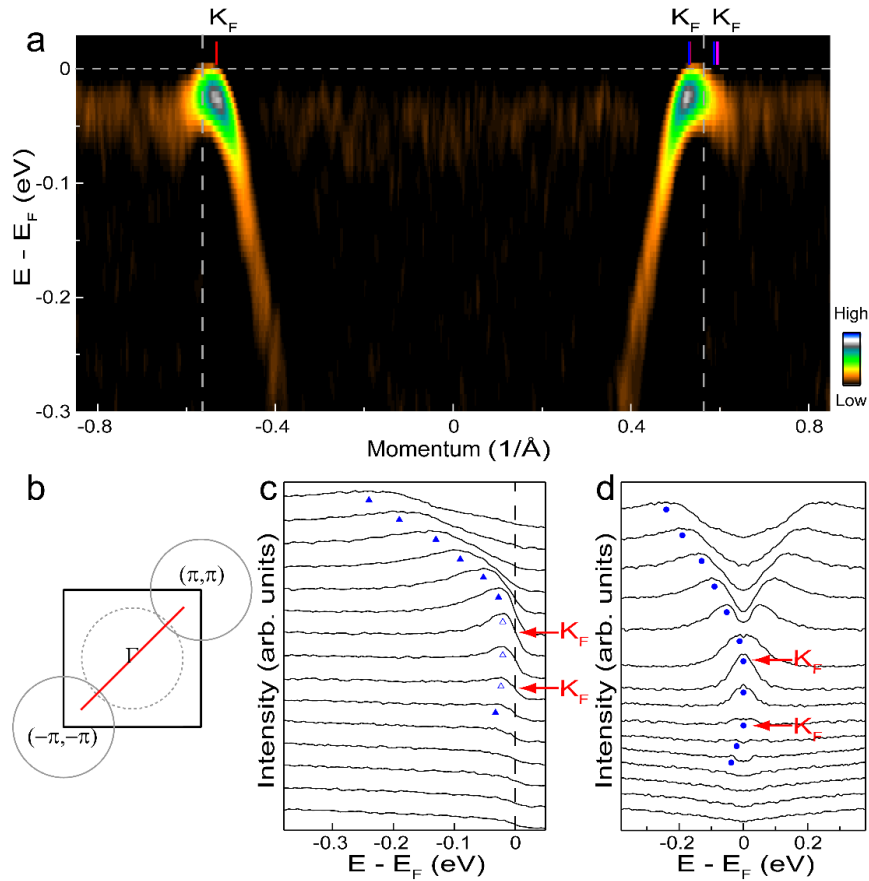

**Fig. 5. Extraction of the hole-like Fermi pocket. a,** Second derivative image with respect to energy measured along the nodal cut. The bars label Fermi momenta. **b,** Momentum location of the cut. **c,** EDCs of the right branch in **a.** The red arrows label the Fermi momenta, where the leading edges of the EDCs cross the Fermi level. **d,** Symmetrized EDCs, which give the same Fermi momenta.

**Extract the Fermi pocket size by the band dispersion and energy gap**

The size of the Fermi pockets can also be estimated by the reconstructed band dispersion and energy gap. If we disregard the origin of the Fermi surface reconstruction, then the two-band model[1-4] (equations 2 and 3) gives a very good phenomenological description of the observed band dispersion. With the measured energy gap, we can calculate the area of the electron (hole) pocket by $\varepsilon_{electron}(\varepsilon_{hole})$.

$$\varepsilon_k = -2t(cosak_x + cosak_y) + 4t'cosak_x\,cosak_y - 2t''(cos2ak_x + cos2ak_y) + \mu \qquad (1)$$

$$\varepsilon_{electron} = \frac{\varepsilon_k + \varepsilon_{k+Q}}{2} + \sqrt{\left(\frac{\varepsilon_k - \varepsilon_{k+Q}}{2}\right)^2 + \Delta^2} \qquad (2)$$

$$\varepsilon_{hole} = \frac{\varepsilon_k + \varepsilon_{k+Q}}{2} - \sqrt{\left(\frac{\varepsilon_k - \varepsilon_{k+Q}}{2}\right)^2 + \Delta^2} \qquad (3)$$

, where $\Delta$ is half of the energy gap between the lower and upper bands, $\varepsilon_k$ is the tight binding bare band dispersion, Q is the scattering wave vector (π/a, π/a), and $\mu$ is the chemical potential determined by the doping level. The parameters in (1) have been determined to be t=0.38eV, t'=0.076eV, t''=0.038eV, which give the best fit to the overall electronic structure of the experiment.

**Reference:**

1. Das T, Markiewicz RS, Bansil A (2006) Nonmonotonic $d_{x^2-y^2}$ superconducting gap in electron-doped $Pr_{0.89}LaCe_{0.11}CuO_4$: evidence of coexisting antiferromagnetism and superconductivity? *Phys Rev B* 74:020506(R).

2. Helm T, et al. (2009) Evolution of the Fermi surface of the electron-doped high-temperature superconductor $Nd_{2-x}Ce_xCuO_4$ revealed by Shubnikov–de Haas oscillations. *Phys Rev Lett* 103:157002.

3. Santander-Syro AF, et al. (2011) Two-fermi-surface superconducting state and a nodal d-wave energy gap of the electron-doped $Sm_{1.85}Ce_{0.15}CuO_{4-\delta}$ cuprate superconductor. *Phys Rev Lett* 106:197002.

4. Armitage NP (2001) Doping the copper-oxygen planes with electrons: the view with photoemission. Ph.D. thesis (Stanford University).

**Weak-coupling approach with short range AFM fluctuation**

Here we show the spectral function simulated by (semi-classical) weak-coupling spin-fluctuation (see, e.g., Refs. 1,2) approach. An AFM correlation length of $\xi \sim 7a$ (over-doped x=0.16 NCCO) [3], is used for the simulation.

As detailed in Ref. 1, the electronic one-loop self-energy $\Sigma_R(\omega, \mathbf{k})$ is dominated in the regime of interest (the "renormalized classical regime") by the zeroth Matsubara frequency and contains the following singular contribution

$$\Sigma_R(\omega, \mathbf{k}) = g \frac{T}{\sqrt{(\omega - \varepsilon_{k+Q})^2 + \Delta^2}} \left[ -i + \frac{1}{\pi} \log \left| \frac{\omega - \varepsilon_{k+Q} + \sqrt{(\omega - \varepsilon_{k+Q})^2 + \Delta^2}}{\omega - \varepsilon_{k+Q} - \sqrt{(\omega - \varepsilon_{k+Q})^2 + \Delta^2}} \right| \right].$$

Here $\omega$ and $\mathbf{k}$ are (real) frequency and momentum, $\varepsilon_k$ is the electronic dispersion, $T$ is temperature, $\mathbf{Q} = (\pi, \pi)^T$ the AFM wavevector, and $\Delta = v_F/\xi$ with Fermi velocity $v_F$. Furthermore, the overall prefactor $g$ is the coupling constant with dimension energy.

As is well-known, this form of the self energy leads to pseudogap-like behavior in the spectral function in the vicinity of the intersection of the bare Fermi surface and the antiferromagnetic zone boundary (AFMZB) if the correlation length is sufficiently large. However, when the coherence length is small, there is neither visible back bending of the band nor clear gap opening at the AFMZB (Fig. 6b). If we simply treat the coupling constant $g$ as an independent fitting parameter and assume an arbitrarily large value, then a gap-like spectral weight suppression may appear, but no back-bending behavior appears at the AFMZB (Fig. 6c). This is different from the experimental data.

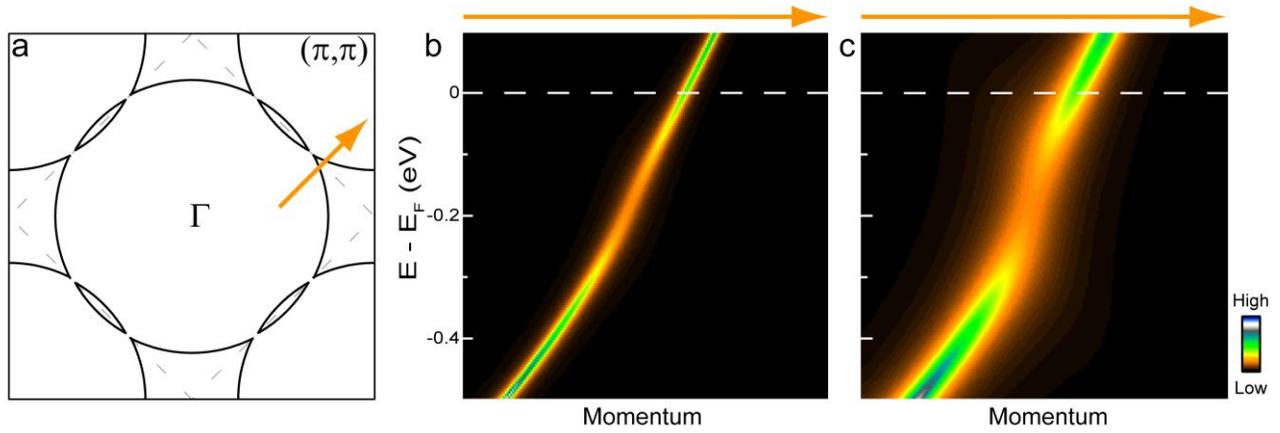

**Fig. 6. Absence of back bending due to small correlation length.** Simulated spectral function along a momentum cut through the electron-like pocket (marked by the arrow in **a**), for different values of $g$, (**b**) $g = 3$ eV (of the order of the bandwidth), (**c**) $g = 10$ eV. We have used the temperature $T = 20$ K of the experiment, and $v_F = 2\sqrt{2}\ ta$ (to estimate the Fermi velocity) together with $\xi = 7a$.

**References:**

1. Vilk YM, Tremblay A-MS (1997) Non-perturbative many-body approach to the Hubbard model and single-particle pseudogap. *J Phys I (France)* 7:1309-1368.
2. Kyung B, Hankevych V, Daré A-M, Tremblay A-MS (2004) Pseudogap and spin fluctuations in the normal state of the electron-doped cuprates. *Phys Rev Lett* 93:147004.
3. Motoyama EM, et al. (2007) Spin correlations in the electron-doped high-transition-temperature superconductor $Nd_{2-x}Ce_xCuO_{4\pm\delta}$. *Nature* 445:186-189.

**Strong coupling picture with short range AFM fluctuation**

In the Hubbard model with strong correlation U, a gap can exist without a long-range AFM order. Fig. 7 shows an example of the single-particle spectral function along a cut near the hotspot, calculated through cluster perturbation theory (CPT) for 15.6% electron doping. Due to the large Coulomb potential, it is no longer possible to treat the interaction as perturbation or mean field. Instead, CPT divides the 2D plain into finite clusters[1]. The spectral function inside the clusters is solved exactly, and the correlation along these boundaries are included through a perturbation. This approach has been shown to correctly capture the single-particle features in the strong correlation limit[2]. The calculation predicts the persistence of AFM gap near the hotspot at the optimal doping with short range fluctuation. The gap opening and back-bending features are visible. This is a direct result of the many-body nature of the entire system, distinct from the weak coupling theory.

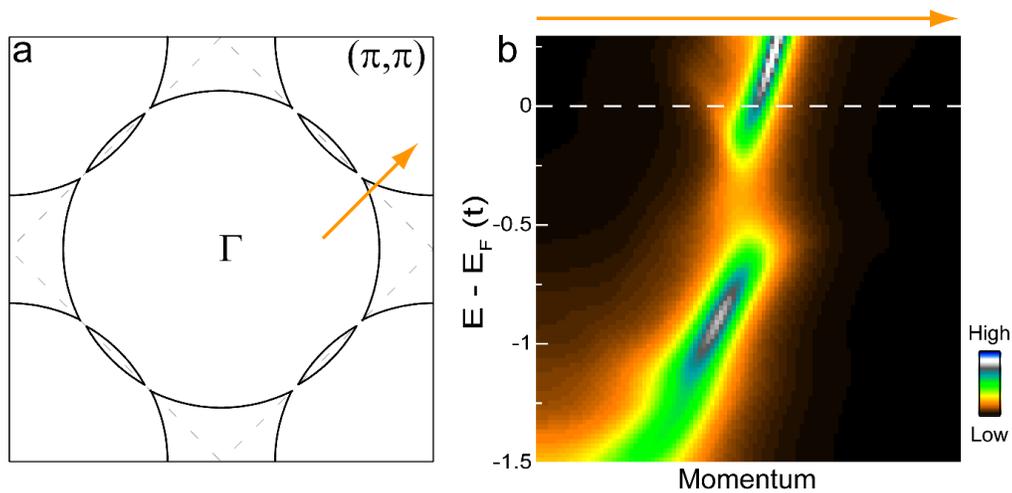

**Fig. 7. Strong coupling calculation.** Simulated spectral function (**b**) along a momentum cut through the electron-like pocket (**a**). The parameters are set as t'=-0.24t, t''= 0.15t and U=6t. The cluster is chosen as 4*4, close to the limit of the state-of-the-art exact diagonalization technique. To achieve the 15.6% doping level, a supercluster technique is adopted in the CPT calculation.

**Reference:**

1. Sénéchal D, Perez D, Pioro-Ladriere M (2000) Spectral weight of the Hubbard model through cluster perturbation theory. *Phys Rev Lett* 84:522.
2. Sénéchal D, Perez D, Plouffe D (2002) Cluster perturbation theory for Hubbard models. *Phys Rev* B 66:075129.

**Spectral function of SU(2) gauge theory with topological order**

Gap opening and back bending of the bands can be obtained in the absence of long-range order and translational symmetry breaking if the system exhibits topological order[1,2]. Fig. 8 shows an example of the calculated spectral function of an SU(2) gauge theory of fluctuating antiferromagnetism within the formalism of Ref. 2 for the parameters relevant to our experiments on NCCO.

The gap opening, band folding and back bending in the simulated spectral function are similar to those observed by ARPES, but there are also additional weak features inside the gap which are not visible in the experimental data.

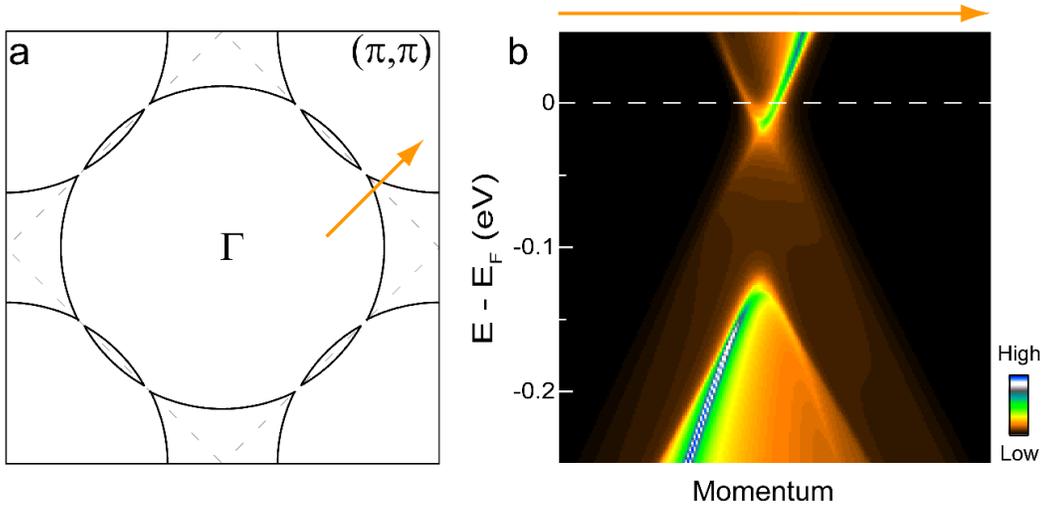

**Fig. 8. Spectrum of SU(2) gauge theory with topological order.** Spectral function $A_\omega(k)$ (**b**) along a momentum cut through the electron-like pocket (**a**). The parameters used in the simulation are: $|H_0| = 0.06t_0$, $T = 0.006t_0$, $\Delta = 0.002t_0$, $t'_0 = -0.45t_0$, $J = 0.1t_0$, $\eta = 0.003\ t_0$, $t = Zt_0 = 380$meV and doping level of $x = 0.15$. Please refer to Ref. 2 for the SU(2) gauge theory and the physical meaning of the parameters.

**Reference:**

1. Sachdev S (2018) Topological order and Fermi surface reconstruction. *arXiv:* 1801.01125v3.
2. Scheurer MS, et al. (2018) Topological order in the pseudogap metal. *Proc Nat Acad Sci* 115:E3665-E3672.